\documentclass[%
reprint,
 amsmath,amssymb,
 aps,
prl,
]{revtex4-1}

\usepackage{soul}
\usepackage{color}

\usepackage{graphicx}
\usepackage{dcolumn}
\usepackage{bm}
\usepackage{hyperref}

\usepackage{xcolor}

\begin{document}
\preprint{APS/123-QED}

\title{Temporal Evolution of Flow in Pore-Networks:\\ From Homogenization to Instability} 

\author{Ahmad Zareei}
\thanks{A.Z. and D.P. contributed equally.}
\affiliation{School of Engineering and Applied Sciences, Harvard University, Cambridge, MA, 02148}
\author{Deng Pan}%
\thanks{A.Z. and D.P. contributed equally.}
\affiliation{School of Engineering and Applied Sciences, Harvard University, Cambridge, MA, 02148}
\author{Ariel Amir}%
\email{arielamir@seas.harvard.edu}
\affiliation{School of Engineering and Applied Sciences, Harvard University, Cambridge, MA, 02148}



\date{\today}

\begin{abstract}
We study the dynamics of flow-networks in porous media using a
pore-network model. First, we consider a class of erosion dynamics
assuming a constitutive law depending on flow rate, local velocities,
or shear stress at the walls. We show that depending on the erosion
law, the flow may become uniform and homogenized or
become unstable and develop channels. By defining an order parameter
capturing these different behaviors we show that a phase transition occurs
depending on the erosion dynamics. Using a simple model, we identify
quantitative criteria to distinguish these regimes and {correctly
predict the fate of the network, and discuss the experimental relevance of our result. }
%


\end{abstract}

%

\maketitle

Fluid flow through a porous medium undergoing a dynamical change in its network of micro-structure is ubiquitous in nature~\cite{marbach2016pruning,alim2013random,tero2010rules,heaton2010growth} as well as in numerous environmental~\cite{schlesinger1999carbon,winkler1979pore,batzle1992seismic} and industrial applications~\cite{duduta2011semi,sun2019hierarchical,smith2017multiphase,ferguson2012nonequilibrium}.
The disordered pore structure of a porous medium results in heterogeneously distributed fluid flow between the pores.
The boundaries of the pore structure can change dynamically either through erosion or deposition/sedimentation of material. Such heterogeneous changes of the solid structure affect the pore-level fluid flow which in turn affects the dynamical changes to the pore structure. This feedback mechanism along with the initial heterogeneous fluid flow makes it difficult to understand and predict the porous media behavior. Nonetheless, an understanding of the dynamical change is essential to improve any of the porous media applications where the pore network changes over time, including groundwater remediation and precipitation of minerals in rocks~\cite{rad2013pore}, biofilm growth in water filtration, and protective filters \cite{herzig1970flow,tien1979advances,jaisi2008transport,carrel2018biofilms,seymour2004anomalous}, as well as enhanced oil recovery with polymer flooding \cite{lake2014fundamentals,parsa2020origin}, or water-driven erosion \cite{schorghofer2004spontaneous,mahadevan2012flow}.

\begin{figure}[h]
    \includegraphics[width = 0.45\textwidth]{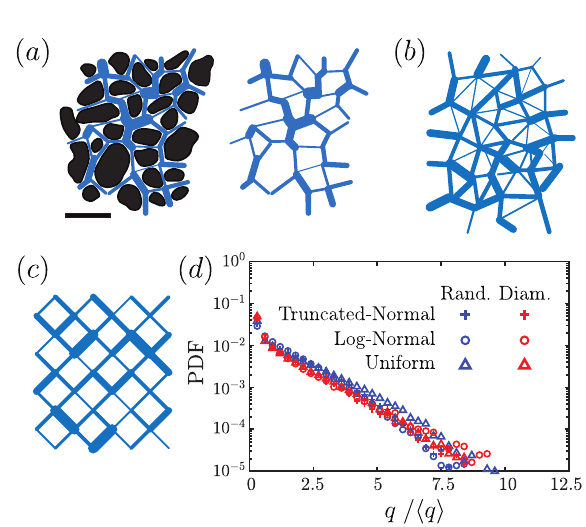}
    \caption{(a) Cross section of a porous sandstone sample obtained using  computerized tomography~\cite{akanji2010finite}. The scale in the bottom left shows $1$mm. The network of pores and throats is highlighted in blue. In the network model, the pores are represented with nodes and the throats between pores are approximated with tubes. (b) Schematic of a topologically random network. The edge diameters representing the pore-throats are randomly distributed. (c) A structured diamond grid network with a random distribution of edge diameters. (d) The universal probability density function (PDF) of fluid flux for a topologically random (blue) or diamond-grid (red) network of nodes (red) with a random distribution of edge diameter sampled from a uniform (triangles), log-normal (circles), or truncated normal (plus) distribution. }    
    \label{fig:fig1}
\end{figure}
\emph{Network approach}-- We approach this long-standing problem using a network model for the porous structure\cite{fatt1956network, blunt2013pore,stoop2019disorder,bryant1993network,blunt2013pore,dong2009pore,blunt2013pore,blunt1995pore,alim2017local}.
The network of pores inside the solid structure is connected together through pore-throats that effectively show resistance to the fluid flow between the pores (Fig.~\ref{fig:fig1}a). Network-based models have been shown to successfully capture key properties of fluid flow in a porous material such as the probability distribution of fluid flux~\cite{alim2017local}, the permeability scaling during clogging~\cite{shima2021}, or the first fluidized path in a porous structure~\cite{fraggedakis_chaparian_tammisola_2021}.  We consider low-Reynolds fluid flow through the porous network. The fluid flow rate at the edge connecting pores $i$ and $j$ is given by $ q_{ij}  = C_{ij}(p_i - p_j)$ where $p_i,p_j$ represent pressures at neighboring nodes. Poiseuille's law implies that the conductance $C_{ij} = {\pi r_{ij}^4}/{8 \mu l_{ij}}$, with $r_{ij}$ and $l_{ij}$ the edge's radius and length. Initially, we consider a topologically random network of nodes constructed using uniformly distributed nodes in a planar domain connected using Delaunay triangulation (Fig. \ref{fig:fig1}b). The pore-throats or radii of the edges are considered as independent and identically distributed random variables and fluid flow in the edges are obtained by solving for conservation of mass at all nodes given a pressure difference between the nodes on the boundaries (supplementary material S1). Independent of edge radius distribution, the probability density function (PDF) of normalized fluid flux is well described by a single exponential distribution shown in Fig. \ref{fig:fig1}d. The exponential distribution of fluid flux is similar to earlier experimental and numerical measurements~\cite{datta2013spatial,shima2021,alim2017local} and is a universal feature in random porous networks. Considering a structured diamond-grid of pores (Fig. \ref{fig:fig1}c) which significantly simplifies the geometrical complexity of the network and allows for analytical derivation, one finds that the PDF of normalized fluid flux remains unchanged for various distributions suggesting robustness to network topology (Fig.~\ref{fig:fig1}d and supplementary materials S2). In the following, we will study, analytically and numerically, how this universal distribution evolves as the network is modified based on a local constitutive law.

\textit{Network evolution}-- The degradation of the solid skeleton (i.e., erosion) or deposition of material on the pore throats (i.e., clogging) in the network of pores is modeled by the change (increase or decrease) in the radii of the edges connecting the pores which translates into changes in the flow resistance between the pores. The rate of change of the radii depends on local fluid flow parameters, however, the exact dependence is unknown. Different models have been used where erosion is assumed to be locally proportional to shear stress at the walls \cite{jager2017channelization,ristroph2012sculpting,hacking1996shear,wan2004investigation}, power dissipation by flow~\cite{steeb2007modeling,marot2012study,sibille2015internal}, or local pressure difference~\cite{derr2020flow,mahadevan2012flow}. We use a general constitutive  model which may implement a diverse set of erosion or clogging dynamics and thus allows us to study the effect of different laws in a unified way. 

\begin{figure}[!h]
    \includegraphics[width = 0.48\textwidth]{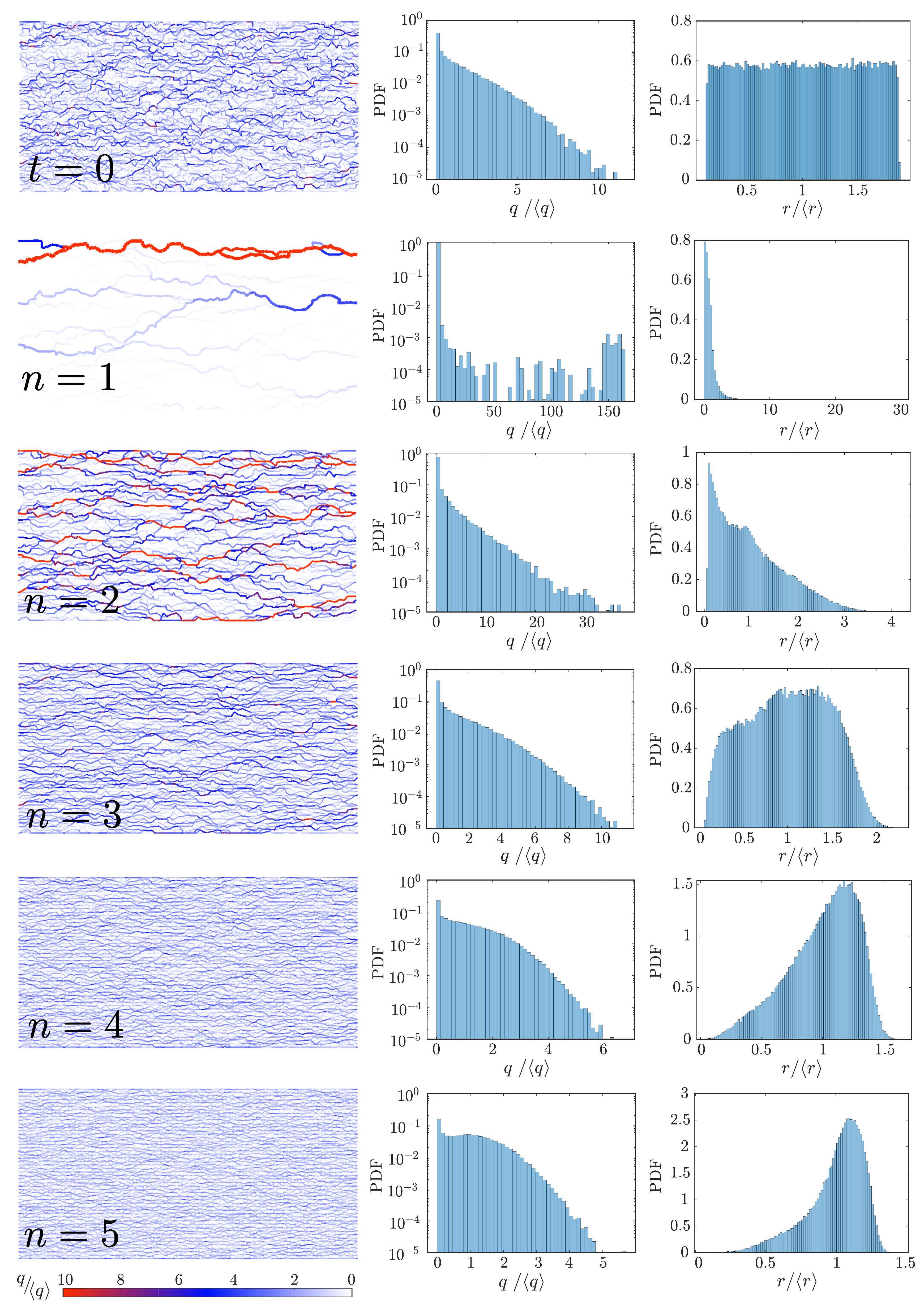}
    \caption{Erosion in a network of pipes. The initial condition is shown with the label $t=0$ in the first row. Each row afterward corresponds to the {simulation result at $t=T$ where  $\langle r_{t=T}\rangle=2r_0$ with $r_0 = \langle r_{t=0}\rangle$}. The erosion law is based on Eq. \eqref{eq:erosion-law}  {where $m=1$ and different powers of $n$ correspond to different models of erosion}. {The first, second, and third columns are snapshots of the pore network, the PDF of normalized fluid flux $q/\langle q \rangle$, and  the PDF of normalized radius $r/\langle r\rangle$ at $t=T$, respectively}.} \label{fig:fig2}
\end{figure}

In order to model erosion in porous media, we consider the abrasion in the throats leading to decrease in the tube radii. We model the dynamics as 
\begin{align}
   \frac{dr_{ij}}{dt} = \alpha  \frac{|q_{ij}|^m}{r_{ij}^n}, \label{eq:erosion-law}
\end{align}
{where $m,n,\alpha$ are constants. Different values of $m$ and $n$ along with a positive $\alpha>0$ correspond to different erosion physics (analogously, $\alpha<0$ corresponds to clogging). Particularly, the erosion when $m=1$ and  (i) $n=0$ depends on the amount of flux $q_{ij}$ passing through the edge; (ii) $n=2$ depends on the local velocities; (iii) $n=3$ depends on the shear force at the boundary of the throat.} {Additionally, $m=2$ and $n=6$ corresponds to models considered in biological transport networks where the radii changes are {proportional to the square of shear stress at the boundary walls}~\cite{hu2013adaptation,ronellenfitsch2016global}.} 
We consider a randomly initialized network with disordered diameters obtained from a uniform distribution (supplementary material S1). 
The flow inside the pores, PDF of flux in the tubes, and PDF of tube radii are shown in Fig. \ref{fig:fig2}. We assume a constant pressure difference between the left and the right boundaries. In each time step, we increase the local radii of the tubes based on the erosion law introduced in Eq. \eqref{eq:erosion-law}, assuming erosion is linear in the flux ($m=1$). Later we will consider the network behavior for other powers of $m$.
We continue the simulations until $\langle r \rangle = 2 r_0$. 
The results of the simulations for different values of $n$ are shown in Fig. \ref{fig:fig2}. When $n=1$ or $2$,  the network develops channels. In such cases, the flow is dominated by a few edges carrying most of the flow while the rest of the network carries almost no flow. This is also reflected by a bimodal radii distribution. In contrast, when $n=3$  (corresponding to erosion linear in shear), we find that despite the 
increase in tube radii and absolute flow rates, the normalized flow distribution is hardly affected, maintaining its original exponential form. Increasing $n$ to larger values, $n=4$ or $5$, we find that the flow pattern in the network moves towards homogenization. Here, the tail of the normalized fluid flux distribution retracts and the coefficient of variation reduces. Similarly, the PDF of the tube diameters becomes narrower. 

{We found that similar results uphold in a 3d random tube network as well as a 2d topologically ordered (diamond grid) network, illustrating the robustness of the results to the network topology (Figs. S2 and S3). Similarly, we found that the results hold also when using an initial narrow distribution of diameters (Fig. S4), showing robustness with respect to the strength of the disorder.}

\begin{figure}[!h]
     \includegraphics[width = 0.45\textwidth]{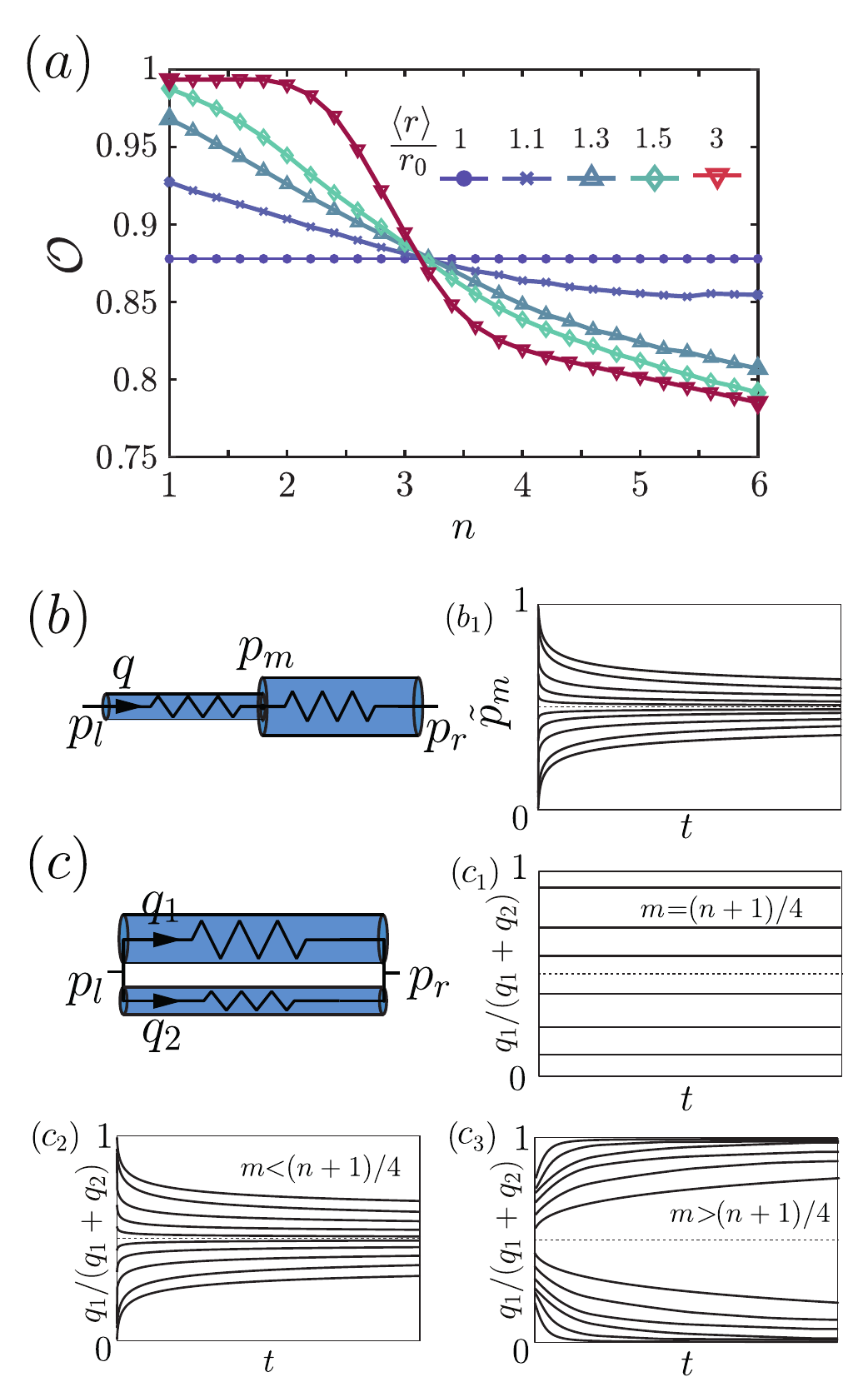}
    \caption{ (a) Order parameter $\mathcal{O}$ calculated from simulation results presented in Fig. \ref{fig:fig2} for different powers of $n$ with $m=1$ plotted over time. (b,c) Tubes in series (b) or parallel (c) configuration. 
    The tube radius dynamically change with the erosion law (Eq. \eqref{eq:erosion-law}). When the tubes are in series, for any $m,n$, (b$_1$) the normalized pressure at the middle junction between the tubes $\tilde p_m = (p_m - p_l)/(p_r-p_l)$ approaches $1/2$ which results in a homogenized pressure distribution. 
    When the tubes are in parallel, for (c$_1$) $m=(n+1)/4$, the flow ratio between the pipes does not change over time; (c$_2$)  $m<(n+1)/4$ the flow distributes between the tubes equally and  $q_1/(q_1+q_2) \to 1$ which results in the homogenization of the network; (c$_3$) $m>(n+1)/4 $, the entire flow eventually passes through one of the tubes, and channeling occurs. }\label{fig:fig3}
\end{figure}

\textit{Phase transition and order parameter}-- {To quantify the transition of the network between the channeling instability and homogenization, we define an order parameter 
\begin{align}
\mathcal{O} = \frac{1}{N-1} \left( N - \frac{\left( \sum_{ij} q^2_{ij}\right)^2}{ \sum_{ij} q^4_{ij}}\right),
\end{align}
where $N$ is the number of edges.} {The order parameter defined here is inspired by the participation ratio (PR) employed to quantify the localization of an eigenstate in the analysis of Anderson localization\cite{kramer1993localization}.}  The order parameter $\mathcal{O}=0 $ when the flux through every edge is identical. On the other hand, when fluid flux becomes highly localized with only a few edges with non-zero flux, $\mathcal{O}\to 1$.  We numerically calculated the  order parameter $\mathcal{O}$ for {randomly initialized networks, averaged over 20 different realization}. 
The results are shown in Fig. \ref{fig:fig3}a for different amounts of erosion measured by the increase in the average diameter $\langle r\rangle /r_0$. As shown in 
 Fig. \ref{fig:fig3}c, at $n\approx 3$ the order parameter, remains unchanged; however for $n>3$ the order parameter moves toward zero, where the flow becomes more uniform, and for $n<3$ the order parameter goes toward unity, where channels are developed. This indicates a phase transition at $n=3$ in the long-time behavior of the network.

\begin{figure}[!h]
    \centering
    \includegraphics[width = 0.45\textwidth]{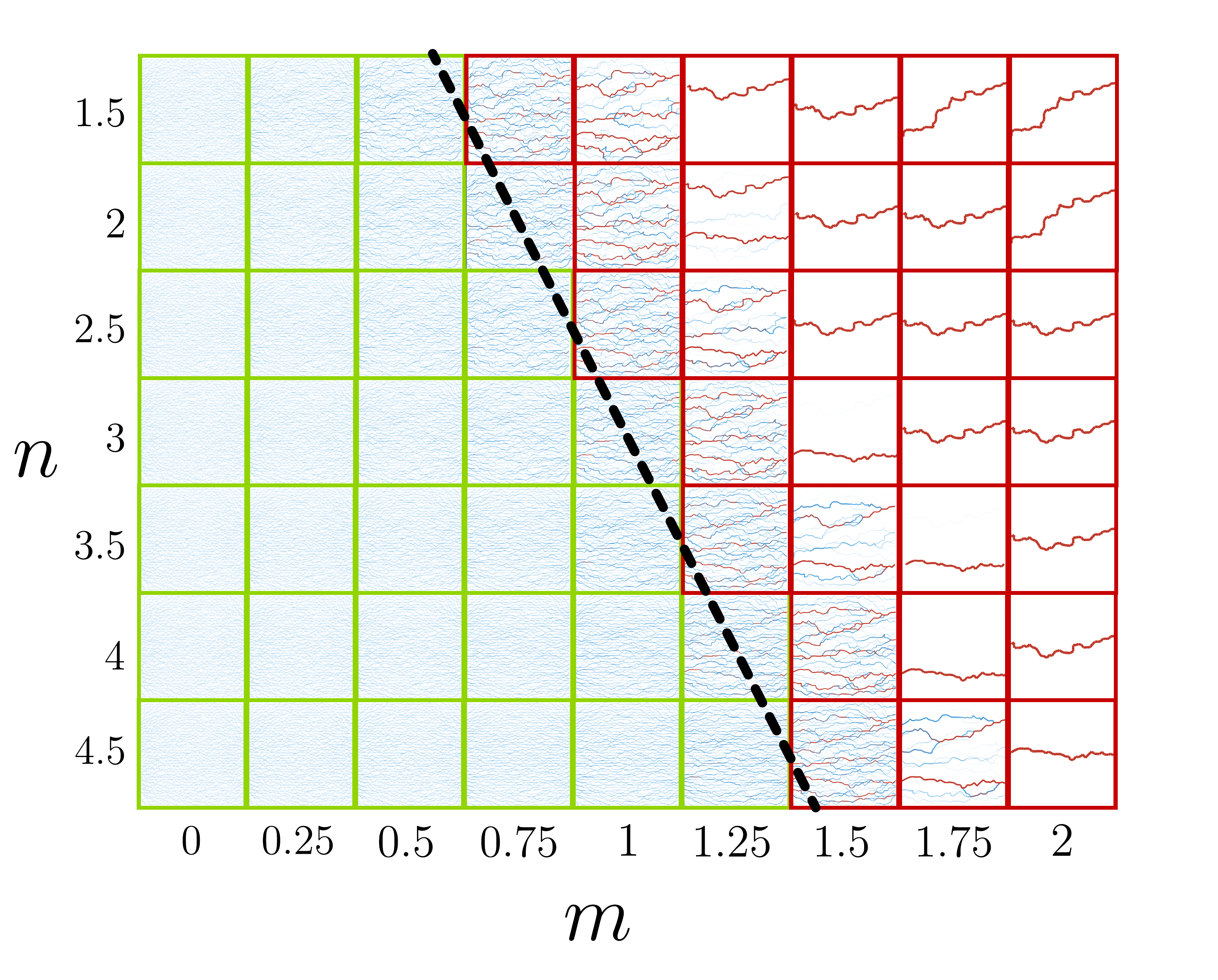}
    \caption{{Evolution of a randomly initialized network for various powers of $m$ and $n$ in Eq. \eqref{eq:erosion-law}}. The network is randomly initialized with $50\times 50$ randomly distributed pores. The bounding box color shows the prediction of the simplified  model for the fate of the network: homogenization (blue) or channelization (red). The dashed black line shows the transition boundary between channelization instability and homogenization obtained using simplified model, i.e., $m=(n+1)/4$.}\label{fig:fig4}
\end{figure}

\textit{Simplified Model}-- {To understand the transition in network behavior during erosion for different powers of $n$, we focus on a simplified model with only two tubes in parallel or series (Figs. \ref{fig:fig3}b-c). First, assuming two cylindrical tubes with radii $r_1,r_2$ in series, the flow is the same for the two tubes $ q_1 = q_2 = q$ (Fig. \ref{fig:fig3}b). The radius of each tube then changes as $dr_i/dt = \alpha q^m /r_i^n$ where $i=1,2$. 
As a result, we find that the conductivity of each tube changes as $dC_i/dt \propto q^m C_i^{(3-n)/4}$, where each tube's conductivity increases. Considering the pressure at the junction between tubes, we find that it moves toward the average value of pressure on both sides (Fig. \ref{fig:fig3}(b$_1$)). Contrary to tubes in series, when the tubes are in parallel (Fig. \ref{fig:fig3}c), the flow divides between the two tubes in proportion to their conductivity, i.e., $q_1/q_2 = C_1/C_2$. Since each tube's radius changes as $dr_i/dt = \alpha q^m_i/r_i^n$, the evolution of the fluid flow ratio becomes 
\begin{align}
\frac{d}{dt}\left( \frac{C_1}{C_2}\right) \propto  \frac{C_1}{C^{n/4+1}_2} \left( \left( \frac{C_1}{C_2}\right)^{m-\frac{n+1}{4}} - 1 \right). \label{eq:dyn-paral}
\end{align}
When $m=(n+1)/4$ in Eq. \eqref{eq:dyn-paral}, the right-hand-side vanishes and as a result the flow ratio $C_1/C_2$ remains constant (Fig. \ref{fig:fig3}c$_1$). However, when $m \neq (n+1)/4$, we find that $C_1/C_2 = 1$ is an equilibrium point. When $m<(n+1)/4$ this equilibrium solution is stable and the flow moves toward homogenization (Fig. \ref{fig:fig3}c$_2$); however, when $m>(n+1)/4$ this equilibrium solution becomes unstable and the solution moves toward $C_1/C_2 \to 0$ or $\infty$ which means that the entire flow passes through one of the tubes (Fig. \ref{fig:fig3}c$_3$). In summary, when the tubes are in series any erosion law makes flow become more uniform; however, when the tubes are in parallel depending on the powers $m,n$ the flow in the tubes can move toward becoming more uniform ($m<(n+1)/4$), maintain the same ratio ($m=(n+1)/4$), or move toward instability and channel development ($m>(n+1)/4$). Since a complex network includes both series and parallel connections, it is plausible that the whole network structure will behave in a similar manner, with a transition in the networks behavior at $ m = (n+1)/4 $.
This observation  is consistent with the numerical simulation results shown in Figs. \ref{fig:fig2} and \ref{fig:fig3}a (for $m=1$) as well as for additional values of $m$ (Fig. \ref{fig:fig4}).}

\textit{Analysis of generalized model}-- So far we focused on erosion dynamics with $m=1$ (Eq. \eqref{eq:erosion-law}) since it directly corresponds to erosion laws of interest, i.e. an erosion rate with a linear dependence to fluid-flux, velocity, or shear-rate at the walls. Considering $m=2$ in Eq. \eqref{eq:erosion-law}, our model aligns with the transport optimization problem in biological networks~\cite{ronellenfitsch2016global,corson2010fluctuations,hu2013adaptation}. Previous works have suggested that in the context of biological transport networks, the network is optimized to minimize its dissipation energy with regards to some constraint (such as constant material or metabolic cost).  Interestingly, the gradient descent method utilized to find the minimal energy configurations maps to Eq. \eqref{eq:erosion-law} with $m=2$, albeit with additional regularizing terms. While under the dynamics we study here erosion will occur indefinitely, in these biological network models a minimal energy configuration exists due to these additional constraints. Nonetheless, the minimal energy configurations manifest a phase transition reminiscent of the one we observe in our model.
To test the role of the parameter $m$ in our model, we simulated the general form of erosion dynamics (Eq. \eqref{eq:erosion-law}).
The simulation results for a randomly initialized network for different powers of $m$ and $n$ are shown in Fig. \ref{fig:fig4}, where each box shows the final snapshot of the network eroded with the corresponding $m$ and $n$.  We further compare the network's simulation result with the prediction of our simplified model for the fate of the network for each pair of $m,n$. The simplified model's prediction is shown using the bounding box color (red for channelization and green for homogenization) in Fig. \ref{fig:fig4} (cf. Fig. S5 in supplementary material showing the heat map for the average change in the order parameter). Additionally, the dashed black line in Fig. \ref{fig:fig4} shows the simplified model's prediction for the boundary between network's transition to homogeneity or channelization (i.e., $m = (n+1)/4$). Although the simplified model is based on the erosion dynamics of two edges in a parallel or series configuration, it still correctly predicts the fate of the network with the complex topology for different values of $m$ and $n$, and captures the boundary separating channelization/homogenization.

In the case of clogging, the initial dynamics can similarly be captured using our model, however, due to the change in the connectivity of the network, our simplified model cannot extend to large time behaviors (supplementary material S5).

\textit{Conclusion}-- {We analyzed the dynamics of porous networks during erosion.  We showed that depending on the form of the erosion law (namely, its dependence on flux and tube radius) the network can either move towards homogenization or towards developing a channeling instability.  We elucidated the physical origin of this phase transition and how it is achieved using a simplified model. Our results highlight the importance of local dynamics and feedback mechanisms in the network's path toward its asymptotic global behavior \cite{ronellenfitsch2016global,corson2010fluctuations,katifori2010damage,hu2013adaptation,ocko2015feedback},  and allow us to infer the local dynamics using large scale observations. As a result, our model can be used as a bulk behavior proxy for determining the local dynamics of erosion in a system~\cite{mahadevan2012flow}. Interestingly, our results indicates that an erosion model that is local and linearly dependent on shear rate cannot result in channelization (since $m=1,n=3$ and $m=(n+1)/4$). However, we note that if the dependence on the shear rate is non-linear, it will qualitatively map to our model albeit with renormalized values of $n$ and $m$: e.g., a power-law dependence on shear rate with exponent $s$ would lead to $m =s$, $n = 3 s$, implying channelization when $s>1$. In the future, it would be exciting to test this and other predictions experimentally on model systems, relying on the technological advances in imaging flow profiles in porous materials, as well as extend the study to the geologically relevant case of chemical erosion \cite{edery2014origins,edery2016characterization}. 
}

\begin{acknowledgments}
We thank the Kavli Foundation, MRSEC  DMR-2011754, and DMR-1420570 for their support. We thank Yaniv Edery,  Eleni Katifori, and Chris Rycroft
for useful discussions.
\end{acknowledgments}



%


\appendix

\newpage
\onecolumngrid

\renewcommand{\thepage}{S\arabic{page}}
\setcounter{page}{1}
\renewcommand{\thesection}{S\arabic{section}}
\setcounter{section}{0}
\renewcommand{\thetable}{S\arabic{table}}
\renewcommand{\thefigure}{S\arabic{figure}}
\setcounter{figure}{0}
\renewcommand{\theequation}{S\arabic{equation}}
\setcounter{equation}{0}
\def\lp{\left(}
\def\rp{\right)}
\def\lb{\left[}
\def\rb{\right]}

{\huge \textbf{Supplementary Material}}

\section{S1. Simulation algorithm}
\label{s1}
In our simulations, we tested two types of networks: (i) a topologically ordered (diamond-grid) network, (ii) a topologically random network (2d and 3d). 
The 2d random network is created using uniformly distributed points with on average $N_x\times N_y$ nodes in the horizontal and vertical directions where the randomly distributed points are connected using a Delaunay triangulation. The 3d random network similarly is obtained by a uniform distribution of $N_x\times N_y \times N_z$ points in space, where the the points are connected using Voronoi cell initialization. The diameter of each edge is sampled from either a uniform distribution with $\mathcal{U}(1,14)$, log-normal distribution with $\mu=3,\sigma=0.48$, or truncated normal distribution with $\mathcal{N}(\mu=7.0,\sigma=3.6)$, where all of the distributions have a coefficient of variation close to $0.5$.  An external pressure is considered between the left-most nodes and the rightmost nodes ($p_\text{left}=10,p_\text{right}=0$). For each edge, assuming a Poiseuille flow, the fluid flux  $q$ and pressure difference $\delta P_e$ are related through $q_e = C_e \delta P_e$, where $C_e = \pi r^4_e/8\mu L_e$,  $L_e$ is the length of the tube, and $\mu$ is the viscosity of the fluid. We define $\vec{q}_e$ as the vector of fluid flux through all the edges, and as a result $\vec{q}_e = \mathbf{C} \mathbf{D}\vec{P}_n$ where $\vec{P}_n$ is the vector of pressure at all the nodes, $\mathbf{D}$ is the transpose of the network's oriented incidence matrix, and $\mathbf{C}$ is the diagonal matrix of edge conductances $\mathbf{C}_e = \text{diag} \left( C^{(1)}_e, C^{(2)}_e, \cdots,C^{(N_e)}_e\right)$. The orientation (or direction) of an edge is arbitrary selected, and it only determines the positive direction for the fluid flow in that edge. Next, we  use conservation of mass at the nodes to solve for the network pressure/flux at the nodes/edges. The conservation of mass at each node is 
\begin{align}
    \vec{q}_n = \mathbf{D}^\top \mathbf{C} \mathbf{D} \vec{P}_n, \label{eq:governeing}
\end{align}
where $\vec{q}_n$ is the vector of total incoming flow to each node. The total incoming flow to an internal node is zero inside the network due to the conservation of mass, and can only be non-zero at the boundary nodes. Without loss of generality, we renumber the boundary nodes to $1,2,\cdots, N_{B}$, where $N_B$ shows the total number of nodes at the boundary. We re-partition Eq. \eqref{eq:governeing} to obtain 

\begin{align}
 \begin{bmatrix} {\mathbf{D}_{b}}^\top \mathbf{C} {\mathbf{D}_{b}}  & \vline &    {\mathbf{D}_{b}}^\top \mathbf{C} {\mathbf{D}_{n}}\\ \hline   
    {\mathbf{D}_{n}}^\top \mathbf{C} {\mathbf{D}_{b}} & \vline & {\mathbf{D}_{n}}^\top \mathbf{C} {\mathbf{D}_{n}}
    \end{bmatrix} \begin{bmatrix} P^{BC}_1 \\ P^{BC}_2 \\ \vdots \\ P_{N_B} \\  \hline  P_{N_B+1} \\ \vdots \\ P_{N_n}
    \end{bmatrix} = \begin{bmatrix} q^{BC}_1 \\ q^{BC}_2 \\ \vdots \\ q^{BC}_{N_B} \\ \hline  0 \\ \vdots \\ 0\end{bmatrix} \to 
    \begin{bmatrix} 
    \mathbf{A}_{bb} & \mathbf{A}_{bn} \\
    \mathbf{A}_{nb} & \mathbf{A}_{nn}       
    \end{bmatrix} \begin{bmatrix} P^{BC}_1 \\ P^{BC}_2 \\ \vdots \\ P^{BC}_{N_B} \\ \hline  P_{N_B+1} \\ \vdots \\ P_{N_n}
    \end{bmatrix} = \begin{bmatrix} q^{BC}_1 \\ q^{BC}_2 \\ \vdots \\ q^{BC}_{N_B}  \\ \hline  0 \\ \vdots \\ 0\end{bmatrix}, 
\end{align}
where $\mathbf{A}_{st} = \mathbf{D}^\top_s \mathbf{C}\mathbf{D}^\top_t$ and  $s,t\in\left\{a,b\right\}$. The first $N_B$ elements of the pressure vector (i.e., $P_1, \cdots, P_{N_B}$) represent the pressure at the boundary nodes and the rest represent the pressure for the internal nodes. In summary the above equations can be written as a combination of two set of linear equations 
\begin{subequations}\label{main-solve-eq}
\begin{align}
  & \mathbf{A}_{bb} \vec{P}_{BC} + \mathbf{A}_{bn}  \vec{P} = \vec{q}_{BC}, \label{maina} \\
& \mathbf{A}_{nb}\vec{P}_{BC} +  \mathbf{A}_{nn} \vec{P} = 0. \label{mainb}
 \end{align}
\end{subequations}
where $\vec{P}_{BC} = [P^{BC}_1, \cdots, P^{BC}_{N_B}]^\top$ is the boundary nodes pressure vector, $\vec{q}_{BC} = [q^{BC}_1, \cdots, q^{BC}_{N_B}]^\top$ is the boundary nodes incoming fluid flux vector, and $\vec{P} = [P_{N_B+1}, \cdots, P_{N_n}]^\top$ is the unknown pressure vector for the rest of the nodes. If the pressure at the boundary is given,  we can use Eq. \eqref{mainb} to solve for the internal pressure values $\vec{P}$, and then use Eq. \eqref{maina} to find the required flux at the boundary nodes $\vec{q}_{BC}$. However, if fluid flux vector at the boundary nodes is given (i.e., $\vec{q}_{BC}$ is known), we need to simultaneously solve  Eqs. \eqref{maina} and \eqref{mainb} to find the boundary pressure vector  $\vec{P}_{BC}$ and internal nodes pressure vector $\vec{P}$. In either case, solving Eq. \eqref{main-solve-eq}  results in the nodes' pressure vector and also the fluid flux vector at the boundary nodes. The fluid flux at each edge can then be calculated using  $\vec{q}_e=\mathbf{C}_e \mathbf{D} \vec{P}_n$. Next, given the flux at each edge  $q_e$, we increase (decrease) the edge radius under erosion (clogging) using 
\begin{align}
    \frac{dr_e}{dt} \propto \pm  \frac{q^m_e}{r_e^n},
\end{align}
and re-iterate the process to solve for the new fluid flux vector. We use a simple forward Euler for time integration. For each iteration, we choose the time step $dt$ so that $\max(dr_e) = 0.1 r_0$, where $r_0$ is the smallest radius among all edges. 
This condition guarantees that at each step a small amount of material is eroded and there is no sudden change in the network. We further test the convergence by decreasing $\max(\Delta r_{ij})$ to half and we observe that the average relative change in the flux vector is $\approx 1.2\%$, and the PDFs remain intact without any notable change. The network size for the 2d/3d networks is $N_x=100,N_y=50$/$N_x=50,N_y=12, N_z=12$ unless mentioned otherwise. In either case, the network has an order of $10^3$ edges. The code is publicly available in a GitHub repository \cite{githubrepo}.

%
%

\section{S2. Analytical results for fluid flux EXP-tail PDF in a diamond network}
\label{s2}
\begin{figure}[h]
  \centering
  \includegraphics[width=.75\textwidth]{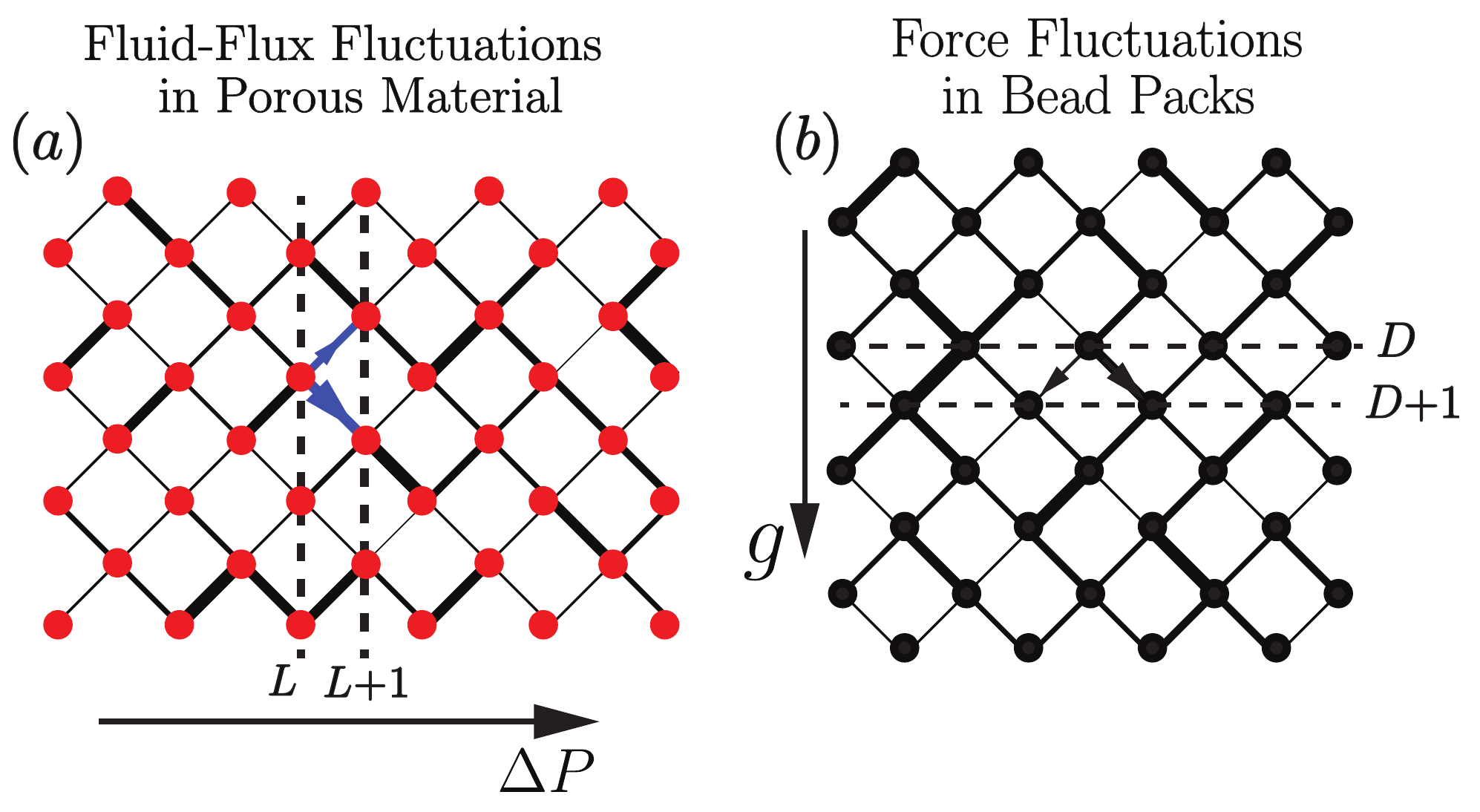} 
  \caption{(a) Schematic of a  diamond grid network of tubes. The incoming flow to each node is redistributed among the outgoing edges. The thickness of the lines shows the fluid flux transferred through that edge. (b) Schematic diagram showing beads (represented with nodes) and their contacts to the neighboring sites (represented with edges). The thickness of the edges show the force transferred through that contact.} \label{grid-result}
\end{figure}

%
%
As described in the main text, the PDF of flow in a topologically disordered network of tubes takes the same form  as in a  structured diamond grid. For completeness, here we repeat the derivation of Refs. \cite{liu1995force,coppersmith1996model,alim2017local} which  show that the observed exponential distribution of fluid flux can be described using a mean-field approach on a structured grid. Basically, the random distribution of the diameters along with the conservation of mass in the network are the two main ingredients resulting in an exponential tail distribution. In a diamond grid, the incoming flow to a node is redistributed among the outgoing edges (since fluid mass is conserved). Due to the randomness in the tube's diameter, the redistribution of the incoming flow to a node between the outgoing edges is random variable.  This model for the flow can be mapped one to one to the problem of force fluctuations in a bead pack \cite{liu1995force,coppersmith1996model,alim2017local} as shown in Fig. \ref{grid-result}. In a bead pack, the force at each layer is redistributed to the next layer where the total force exerted on the next layer should equal to that of the previous layer. 
Given the above conditions, the flow at layer $L+1$ at node $j$ can be obtained as 
\begin{align}
  q(L+1,j) = \sum_i w_{ij} q(L,i) = w_{i,i+1} q(L,i+1) + w_{i,i} q(L,i), \label{total-flow}
\end{align}
where $w_{ij}$ shows the weights by which the flow is redistributed, and $\sum_{j} w_{ij} = 1$ since the total fluid flux is conserved. 
%
%
Assuming a general distribution of for the weights, $\eta(w)$, we can use a mean-field approximation to find the distribution of $q$ at the layer $L$, i.e., $p_L(q)$, as 
\begin{align}
  p_L (q) = \prod_{j=1}^N \left\{ \int_0^1 d w_j \eta(w_j) \int_0^{\infty} dq_j p_{L-1}(q_j)\right\} \times \delta \left( \sum_j w_j q_{j} - q \right),
\end{align}
%
where $N$ is the number of outgoing edges (e.g., in our structured diamond grid $N=2$) and $\delta(\cdot)$ is the Dirac delta function. 
Taking the Laplace transform of the above equation and defining $\tilde p(s) \equiv \int_0^{\infty} p(q) e^{-qs} dq$ one obtains
%
%
%
\begin{align}
  {\tilde P_L(s) = \lp  \int_0^1 d w \eta(w) \tilde P_{L-1}(s w)  \rp^{N}}. \label{eq:main-laplace}
\end{align}
The above equation gives a recursive relation for the Laplace transformed of fluid flux PDF, $\tilde P_L(s)$, where it gradually converges to a distribution $\tilde P(s)$ from which the PDF of fluid flux can be obtained. Solving the above equation for a structured diamond grid network, one finds that the converged PDF of the fluid flux becomes $p(q) = 4q\exp(-2q)$ \cite{liu1995force,coppersmith1996model,alim2017local}, which is an exponential tailed distribution.


\newpage
\section{S3. Robustness to topology and initial condition}
\label{s3}
\begin{figure}[htp]
      \includegraphics[width = 1.0\textwidth]{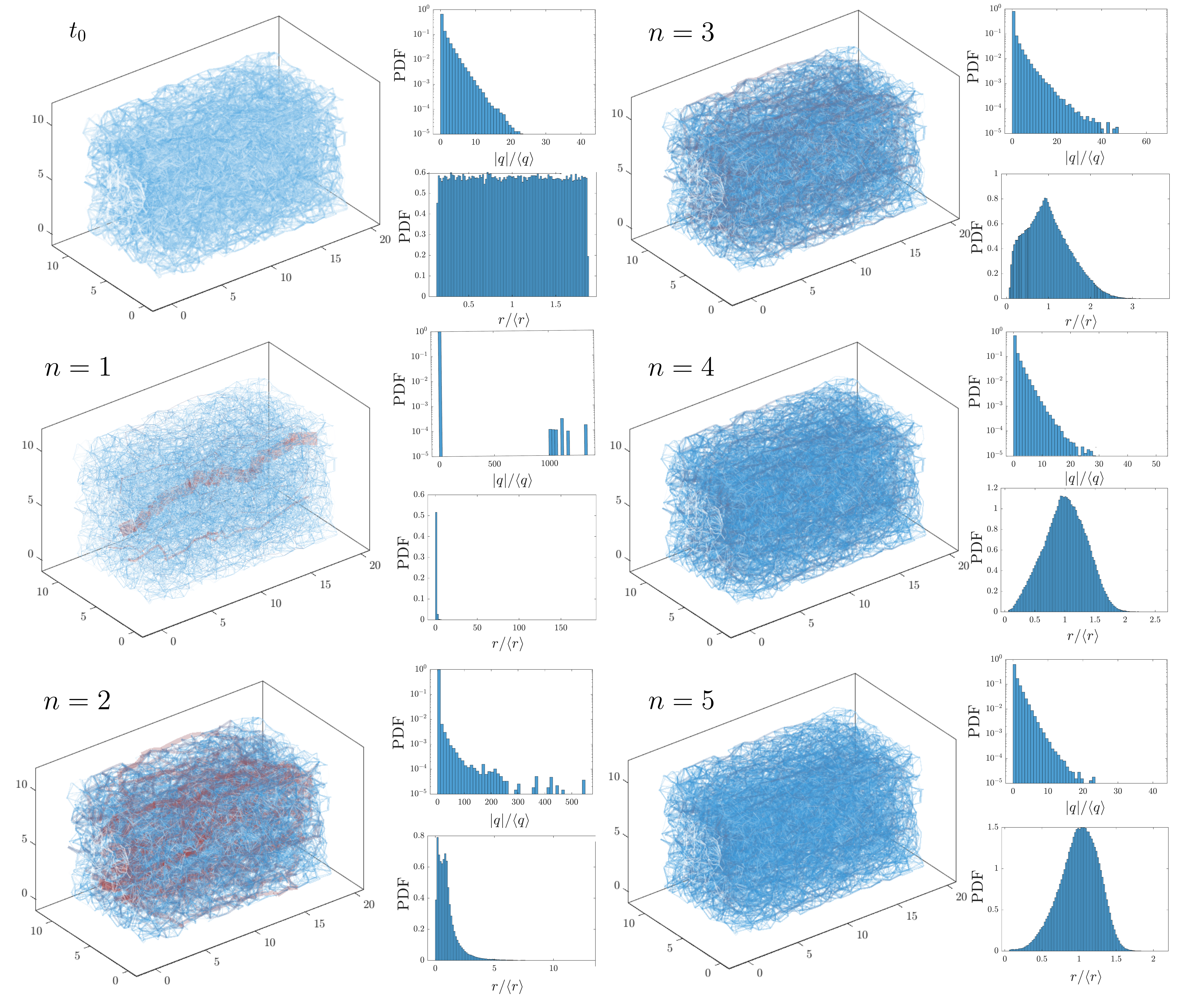}
     \caption{Erosion in a topologically random 3D network of tubes with $N_x=20, N_y=12, N_z=12$ points for Voronoi cell initialization and an initial uniform broad distribution of tube diameters randomly sampled from $\mathcal{U}(1,14)$. Snapshots of the network, PDF of normalized fluid flux $q/\langle q \rangle$, and normalized edge radius distribution $r/\langle r \rangle$ at the initial time $t=0$, and also after $N$ erosion steps for different powers of erosion $n$ are shown. We stop the erosion after $N$ steps such that $\langle r\rangle=2r_0$ where $r_0 = \langle r_{t=0}\rangle$. The erosion law is based on Eq. (1) in the main text where different powers of $n$ correspond to different models of erosion.}\label{SIfig:fig2-3d}
 \end{figure}

\begin{figure}[htp]
 \includegraphics[width=0.75\textwidth]{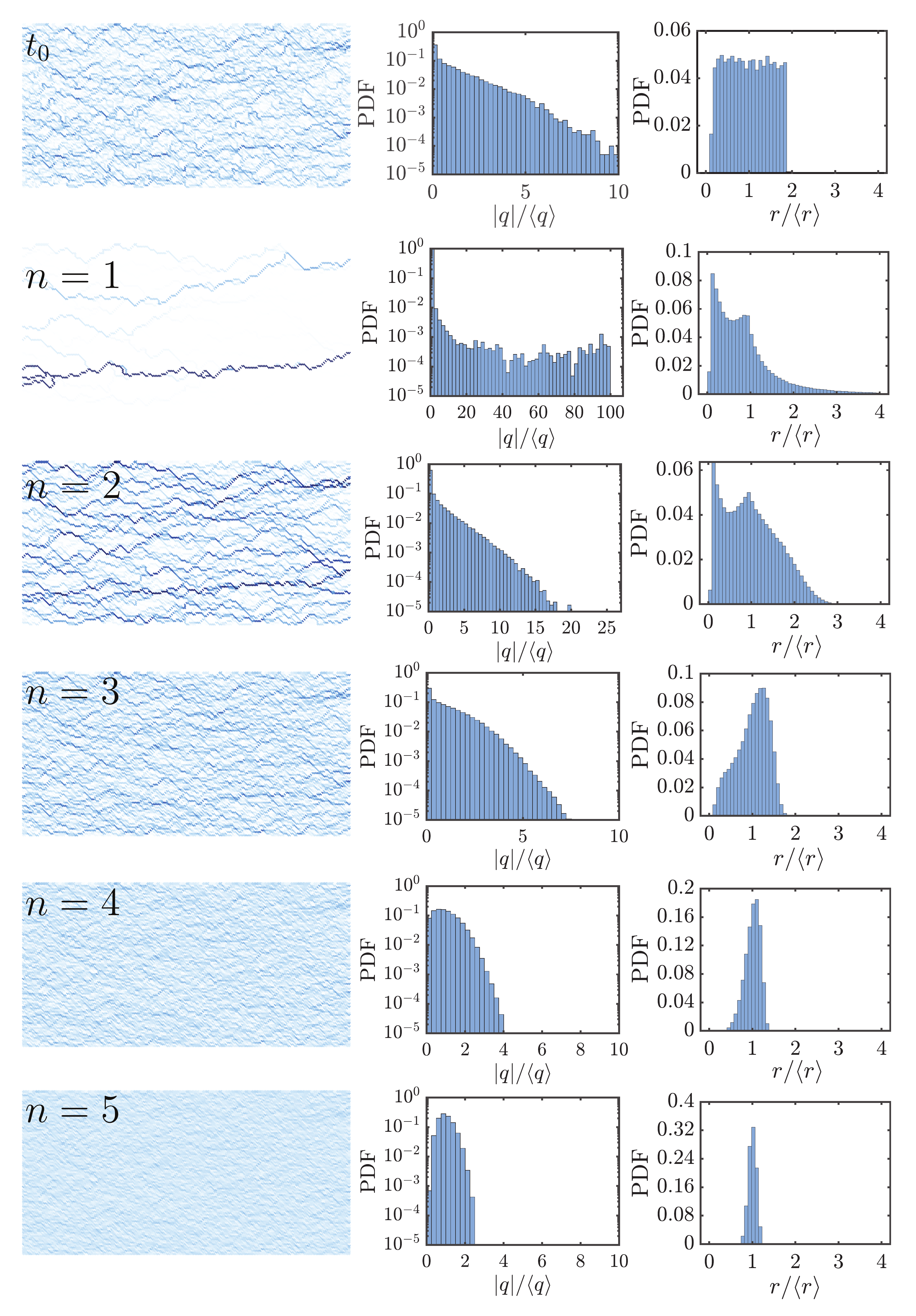} 
     \caption{Erosion in a diamond grid network with $N_x=100, N_y=50$ randomly distributed nodes and an initially broad  distribution of tube diameter randomly sampled from $\mathcal{U}(1,14)$. The initial condition is shown with the label $t=0$ in the first row. Each row afterward corresponds to the simulation result after $N$ steps such that $\langle r\rangle=2r_0$ where $r_0 = \langle r_{t=0}\rangle$ or twice the initial average radius. The erosion law is based on Eq. (1) in the main text where different powers of $n$ correspond to different models of erosion. The first column is a snapshot of the pore network, the second column is the PDF of normalized fluid flux $q/\langle q \rangle$, and the last column is the PDF of normalized radius $r/\langle r\rangle$.}\label{fig:fig2-diamond}%
\end{figure}

\begin{figure}[htp]
 \includegraphics[width=0.72\textwidth]{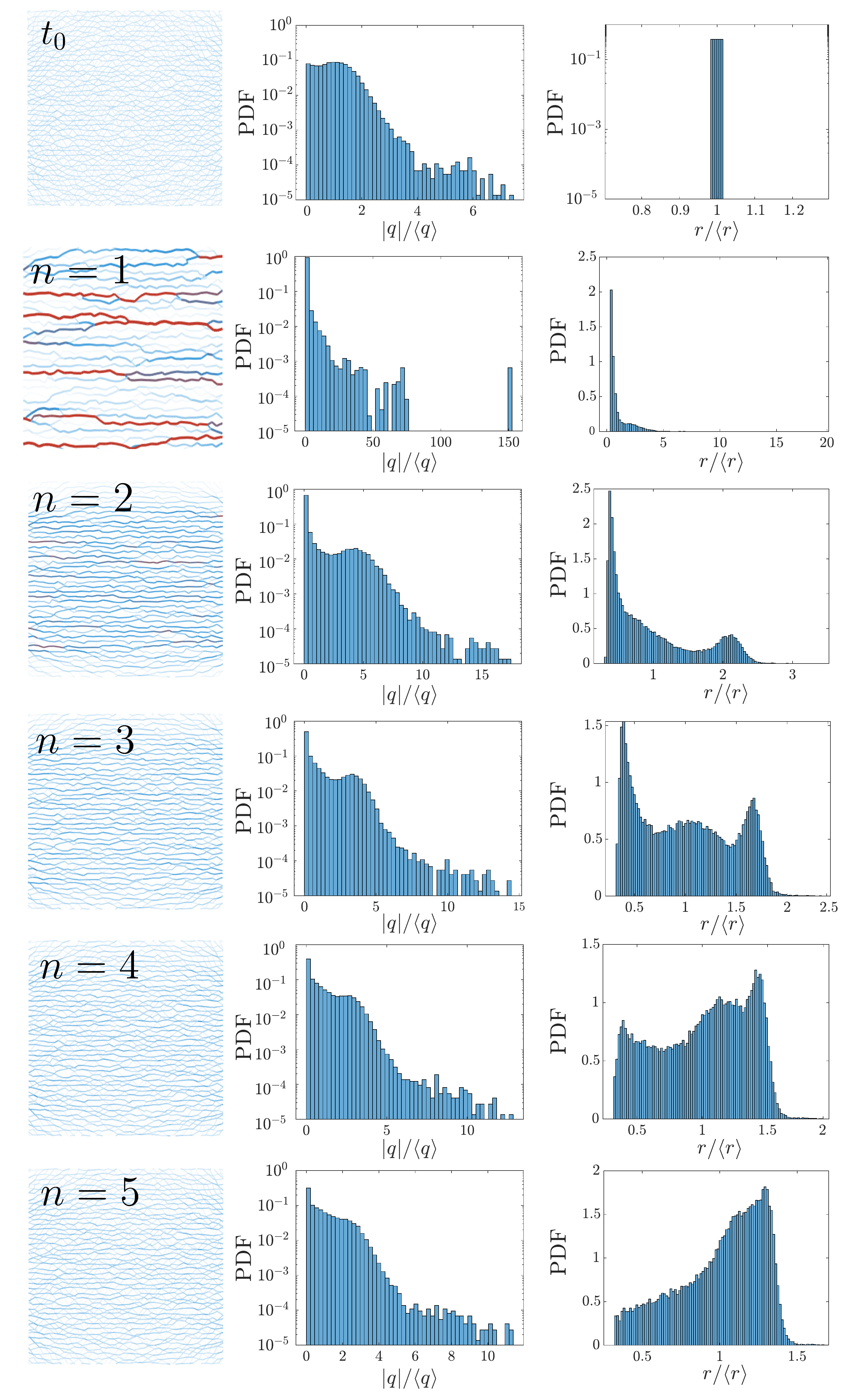}
     \caption{Erosion in a structured diamond grid network with $N_x=50, N_y=50$ and an initially narrow distribution of tube diameters randomly sampled from  $\mathcal{U}\left(d_0(1-\epsilon), d_u(1+\epsilon)\right) $ where  $d_0$ is the average diameter and $\epsilon=0.03$. The initial condition is shown with the label $t=0$ in the first row. Each row afterward corresponds to the simulation result after $N$ steps such that $\langle r_{t=N}\rangle=2r_0$ where $r_0 = \langle r_{t=0}\rangle$ or twice the initial average radius. The erosion law is based on Eq. (1) in the main text where different powers of $n$ correspond to different models of erosion. The first column is a snapshot of the pore network, the second column is the PDF of normalized fluid flux $q/\langle q \rangle$, and the last column is the PDF of normalized radius $r/\langle r\rangle$.}\label{SI_figs4:fig:fig2}%
\end{figure}

In order to check the robustness of our result with respect to the topology of the network, we run our simulations on a three-dimensional random network with Voronoi cell initialization of nodes in space (Fig. \ref{SIfig:fig2-3d}), and also for a two dimensional topologically ordered network with a diamond grid (Fig. \ref{fig:fig2-diamond}), where  both networks are initialized with a uniform random distribution for the diameter of the tubes. We find that regardless of the topology of the network, an erosion dynamics with $n>3$ results in a homogenized network, while $n<3$ results in the channelization instability. We further check the effect of initial randomness on the fate of the network. We use a two-dimensional diamond-grid network with a narrow uniform initial distribution of tube diameters around an average diameter $d_0$ with only a very small variation ($3\%$), i.e., the tube diameters are sampled from $\mathcal{U}(d_0(1-\epsilon), d_0(1+\epsilon))$. We again find that for the networks with $n<3$ channels are formed, while for $n>3$ the network stays homogenized (Fig. \ref{SI_figs4:fig:fig2}).  

\newpage 
\newpage 
\newpage
\section{S4. Average Change in the  order parameter}
\label{s4}
\begin{figure}[htp]
    \includegraphics[width = 0.70\textwidth]{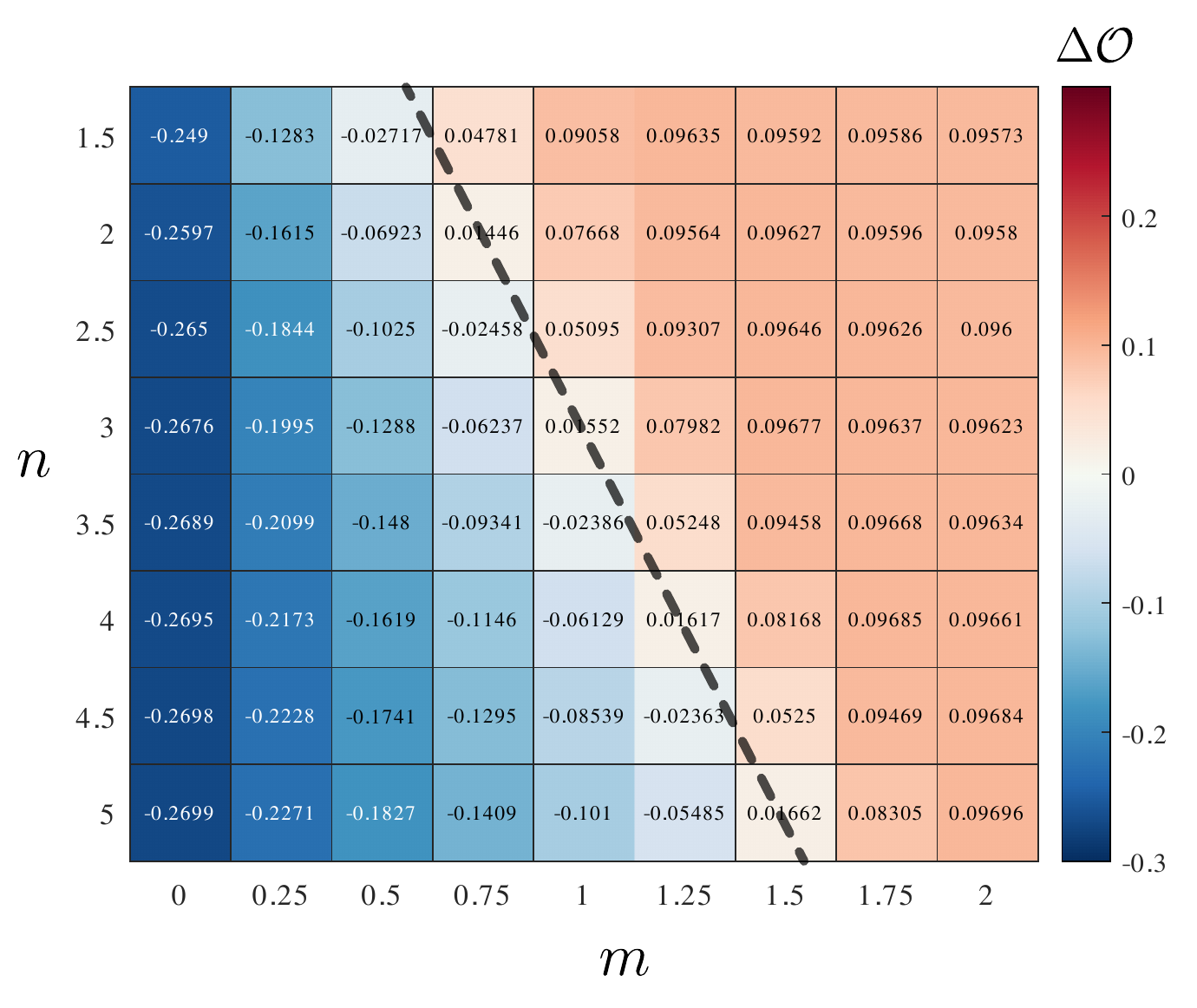}
    \caption{The heat map for the average change in the order parameter corresponding to the networks shown in Fig. 4 of the main text. The order parameter shown in the heat map here is an average of 100 different simulations with different random initialization of tube diameters. The black line shows the boundary between two phases calculated using the model introduced in the main text.}\label{fig:fig4_SI-2}
\end{figure}

In order to quantify the network behavior shown in Fig. 4, we calculate the change in the order parameter for different $m,n$ averaged over 100 simulations with different random initial conditions, and the heat-map results are shown in Fig.~\ref{fig:fig4_SI-2}. The positive or negative change in the order parameter shows the network's change toward homogenization or channelization. The boundary between the two phases (homogenization and channelization) calculated using the simple model introduced in the main text is shown with a solid black line here, and it can be seen that it agrees well with the order parameter change. 

\section{S5. Clogging dynamics}
\label{s5}
%
Besides erosion, another change in the network is the deposition/sedimentation of material on the boundary walls of the porous material. We refer to this dynamical change a ``clogging'' process as opposed to erosion. Contrary to erosion, the clogging behavior may cause some edges to block which effectively alters the network of connectivity and network behavior. This change in the connection between nodes through edges getting blocked can drastically alter porous structure behavior, e.g., causes a huge difference between effective and true porosity \cite{shima2021}. Despite the drastic change of network with blockages, we can still focus on the \textit{initial} change in the order parameter. The derivative of order parameter can be written as 
\begin{align}
    \frac{d\mathcal{O}}{dt} = \sum_{ij} \sum_{kl} \frac{\partial \mathcal{O}}{\partial q_{ij}} \frac{\partial q_{ij}}{\partial C_{kl}}  \frac{\partial C_{kl}}{\partial t} \label{eq:order-derivative}
\end{align}
where the last term changes sign from erosion to clogging, i.e., $\partial C_{kl}/\partial t = \pm \alpha \pi q^m_{kl} /r_{kl}^{n-3}\mu l_{kl}$ for erosion and clogging respectively. As a result, the magnitude of change in the order parameter equals that of erosion. Note that in Eq. \eqref{eq:order-derivative}, the second term depends on the network topology, and pore throat clogging results in the change of network topology at later times.
At short times, however, similar to the erosion, a phase transition exists at $n=3$. When  $n<3$ the network moves toward homogenization during the clogging process and when $n>3$ the flow moves toward the development of channeling instability. At later times, this initial trend, however, might not hold true due to the aforementioned complex changes in the connectivity network during the clogging process.


\end{document}